\documentclass[final,prl,nofootinbib,twocolumn,showpacs,showkeys,groupaddress,preprintnumbers,floatfix]{revtex4-1}

\usepackage{dynlearn}

\definecolor{my_purple}{RGB}{127, 0, 255}

\colorlet{varcolor} {green!50!black}
\colorlet{dycolor}  {red!50!black}
\colorlet{tricolor} {blue!50!black}
\colorlet{opcolor}  {my_purple!50!black}

\colorlet{rowcolor} {gray!25!white}

\newcommand{\var}[1]{\textcolor{varcolor}{\ensuremath{#1}}}
\newcommand{\dy} [1]{\textcolor{dycolor}{#1}}
\newcommand{\tri}[1]{\textcolor{tricolor}{#1}}

\newcommand{\xor}{\textsc{xor}\xspace}

\newcommand{\varsa}{\var{X},\var{Y},\var{Z}}
\newcommand{\varsb}{\var{X}:\var{Y}:\var{Z}}
\newcommand{\varsc}{\var{X}:\var{Y}\downarrow\var{Z}}
\newcommand{\varsd}{\var{X}:\var{Y}\Downarrow\var{Z}}

\begin{document}

\def\ourTitle{%
  Multivariate Dependence Beyond Shannon Information
}

\def\ourAbstract{%
  Accurately determining dependency structure is critical to discovering a system's causal organization. We recently showed that the transfer entropy fails in a key aspect of this---measuring information flow---due to its conflation of dyadic and polyadic relationships. We extend this observation to demonstrate that this is true of all such Shannon information measures when used to analyze multivariate dependencies. This has broad implications, particularly when employing information to express the organization and mechanisms embedded in complex systems, including the burgeoning efforts to combine complex network theory with information theory. Here, we do not suggest that any aspect of information theory is \emph{wrong}. Rather, the vast majority of its informational measures are simply inadequate for determining the meaningful dependency structure within joint probability distributions. Therefore, such information measures are inadequate for discovering intrinsic causal relations. We close by demonstrating that such distributions exist across an arbitrary set of variables.
}

\def\ourKeywords{%
   stochastic process, transfer entropy, causation entropy,
   partial information decomposition, network science, causality
}

\hypersetup{
  pdfauthor={Ryan G. James},
  pdftitle={\ourTitle},
  pdfsubject={\ourAbstract},
  pdfproducer={},
  pdfcreator={}
}

\author{Ryan G. James}
\email{rgjames@ucdavis.edu}
\affiliation{Complexity Sciences Center}
\affiliation{Physics Department\\
University of California at Davis, One Shields Avenue, Davis, CA 95616}

\author{James P. Crutchfield}
\email{chaos@ucdavis.edu}
\affiliation{Complexity Sciences Center}
\affiliation{Physics Department\\
University of California at Davis, One Shields Avenue, Davis, CA 95616}

\date{\today}
\bibliographystyle{unsrt}

\title{\ourTitle}

\begin{abstract}

\ourAbstract

\vspace{0.1in}
\noindent

\end{abstract}


\preprint{\sfiwp{16-09-XXX}}
\preprint{\arxiv{1609.01233}}

\title{\ourTitle}
\date{\today}
\maketitle

\setstretch{1.1}

\section{Introduction}
\label{sec:introduction}

Information theory is a general, broadly applicable framework for understanding a system's statistical properties~\cite{Kull68}. Due to its abstraction from details of the underlying system coordinates and its focus on probability distributions, it has found many successes outside of its original domain of communication in the physical, biological, and social sciences~\cite{Quas53a,Quas58a,kelly1956new,Bril62a,bialek1991reading,strong1998entropy,ulanowicz2011central,grandy2008entropy,harte2011maximum,nalewajski2006information,garland2014model,kafri2014information,varn2015chaotic,varn2016did,zhou2016information,kirst2016dynamic,izquierdo2015information,james2014chaos}. Often, the issue on which it is brought to bear is discovering and quantifying dependencies or causal relations~\cite{schreiber2000measuring,fiedor2014partial,sun2014causation,lizier2008local,walker2015informational,lee2015assessing}.

The past two decades, however, produced a small but important body of results detailing how standard Shannon information measures are unsatisfactory for determining some aspects of dependency and shared information. Within information-theoretic cryptography, the conditional mutual information has proven to be a poor bound on secret key agreement~\cite{maurer1997intrinsic,renner2003new}. The conditional mutual information has also been shown to be unable to accurately measure information flow~\cite[and references therein]{PhysRevLett.116.238701}. Finally, the inability of standard methods of decomposing the joint entropy to provide any semantic understanding of how information is shared has motivated entirely new methods of decomposing information~\cite{williams2010nonnegative,bertschinger2013shared}. Common to all these is the fact that conditional mutual information conflates intrinsic dependence with conditional dependence.

Here, we demonstrate a related, but deeper issue: Shannon information measures---entropy, mutual information, and their conditional and multivariate versions---can fail to distinguish joint distributions with vastly differing internal dependencies.

Concretely, we start by constructing two joint distributions, one with dyadic subdependencies and the other with strictly triadic subdependencies. From there, we demonstrate that no standard Shannon-like information measure, and exceedingly few nonstandard methods, can distinguish the two. Stately plainly: when viewed through Shannon's lens, these two distributions are erroneously equivalent. While distinguishing these two (and their internal causality) may not be relevant to a mathematical theory of communication, it is absolutely critical to a mathematical theory of information storage, transfer, and modification~\cite{lizier2010local}. We then demonstrate two ways in which these failures generalize to the multivariate case. The first generalizes our two distributions to the multivariate and polyadic case via ``dyadic camouflage''. And, the second details a method of embedding an arbitrary distribution into a larger variable space using hierarchical dependencies, a technique we term ``dependency diffusion''. In this way, one sees that the initial concerns about information measures can arise in virtually any statistical multivariate analysis.

In this short development, we assume a working knowledge of information theory, such as found in standard textbooks~\cite{cover2012elements,yeung2012first,csiszar2011information,mackay2003information}.

\section{Development}
\label{sec:development}

We begin by considering the two joint distributions shown in Table\nobreakspace \ref {tab:distributions}. The first represents \emph{dyadic} relationships between three random variables $X$, $Y$, and $Z$. And, the second \emph{triadic}\footnote{This distribution was first considered in Ref.~\cite{griffith2014quantifying}, though for other reasons.} between them. These appellations are used for reasons that will soon be apparent. How are these distributions structured? Are they structured identically, or are they qualitatively distinct?

\begin{figure}
  \centering
  \subfloat[Dyadic]
  {
    \centering
    \includegraphics{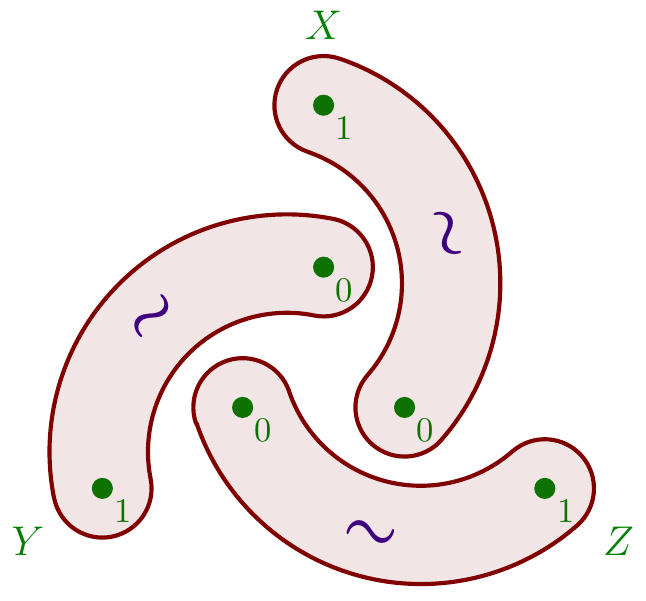}
    \label{subfig:dyadic_dependencies}
  }\\
  \subfloat[Triadic]
  {
    \centering
    \includegraphics{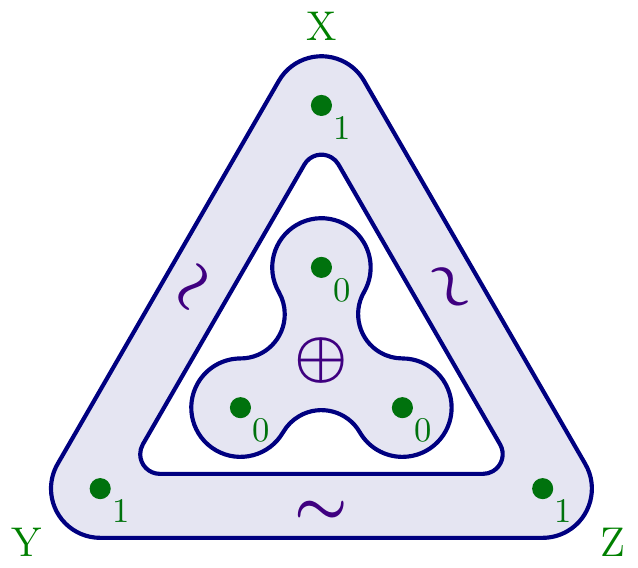}
    \label{subfig:triadic_dependencies}
  }
  \caption{
    Dependency structure for the (a) dyadic and (b) triadic distributions. Here, $\sim$ denotes that two or more variables are distributed identically and $\oplus$ denotes the enclosed variables form the \xor relation. Note that although these dependency structures are fundamentally distinct, their Bayesian network (Fig.\nobreakspace \ref {fig:dependencies}) and their information diagrams (Fig.\nobreakspace \ref {fig:idiagrams}) are identical.
  }
  \label{fig:dependencies}
\end{figure}

We can develop a direct picture of underlying dependency structure by casting the random variables' four-symbol alphabet used in Table\nobreakspace \ref {tab:distributions} into composite binary random variables, as displayed in Table\nobreakspace \ref {tab:binary}. It can be readily verified that the dyadic distribution follows three simple rules: $X_0 = Y_1$, $Y_0 = Z_1$, and $Z_0 = X_1$. Three dyadic rules. The triadic distribution similarly follows simple rules: $X_0 + Y_0 + Z_0 = 0 \mod 2$ (the \xor relation~\cite{cook2005networks}, or, equivalently, any one of them is the \xor of the other two), and $X_1 = Y_1 = Z_1$. Two triadic rules. These dependency structures are represented pictorially in Fig.\nobreakspace \ref {fig:dependencies}. Our development from this point on will not use any knowledge of this structure, but rather it will attempt to determine the structure using only information measures.

\begin{table}
  \centering
  \subfloat[Dyadic]
  {
    \centering
    \begin{tabular}{ccccccl}
      \toprule
      \var{X} && \var{Y} && \var{Z} && $\Prob$ \\
      \midrule
      \dy{0} && \dy{0} && \dy{0} && \nicefrac{1}{8} \\
      \dy{0} && \dy{2} && \dy{1} && \nicefrac{1}{8} \\
      \dy{1} && \dy{0} && \dy{2} && \nicefrac{1}{8} \\
      \dy{1} && \dy{2} && \dy{3} && \nicefrac{1}{8} \\
      \dy{2} && \dy{1} && \dy{0} && \nicefrac{1}{8} \\
      \dy{2} && \dy{3} && \dy{1} && \nicefrac{1}{8} \\
      \dy{3} && \dy{1} && \dy{2} && \nicefrac{1}{8} \\
      \dy{3} && \dy{3} && \dy{3} && \nicefrac{1}{8} \\
      \bottomrule
    \end{tabular}
    \label{subtab:dyadic}
  }
  \quad\quad\quad
  \subfloat[Triadic]
  {
    \centering
    \begin{tabular}{ccccccl}
      \toprule
      \var{X} && \var{Y} && \var{Z} && $\Prob$ \\
      \midrule
      \tri{0} && \tri{0} && \tri{0} && \nicefrac{1}{8} \\
      \tri{1} && \tri{1} && \tri{1} && \nicefrac{1}{8} \\
      \tri{0} && \tri{2} && \tri{2} && \nicefrac{1}{8} \\
      \tri{1} && \tri{3} && \tri{3} && \nicefrac{1}{8} \\
      \tri{2} && \tri{0} && \tri{2} && \nicefrac{1}{8} \\
      \tri{3} && \tri{1} && \tri{3} && \nicefrac{1}{8} \\
      \tri{2} && \tri{2} && \tri{0} && \nicefrac{1}{8} \\
      \tri{3} && \tri{3} && \tri{1} && \nicefrac{1}{8} \\
      \bottomrule
    \end{tabular}
    \label{subtab:triadic}
  }
  \caption{
    The (a) dyadic and (b) triadic probability distributions over the three random variables \var{X}, \var{Y}, and \var{Z} that take values in the four-letter alphabet $\{0, 1, 2, 3\}$. Though not directly apparent from their tables of joint probabilities, the dyadic distribution is built from dyadic (pairwise) subdependencies while the triadic from triadic (three-way) subdependencies.
  }
  \label{tab:distributions}
\end{table}

\begin{table}
  \centering
  \subfloat[Dyadic]
  {
    \begin{tabular}{@{}cccccccccl@{}}
      \toprule
      \multicolumn{2}{c}{\var{X}} && \multicolumn{2}{c}{\var{Y}} && \multicolumn{2}{c}{\var{Z}} && \\
      \cmidrule(lr){1-2}  \cmidrule(lr){4-5}  \cmidrule(lr){7-8}
      \var{X_0} & \var{X_1} && \var{Y_0} & \var{Y_1} && \var{Z_0} & \var{Z_1} && $\Prob$ \\
      \midrule
      \dy{0} & \dy{0} && \dy{0} & \dy{0} && \dy{0} & \dy{0} && \nicefrac{1}{8} \\
      \dy{0} & \dy{0} && \dy{1} & \dy{0} && \dy{0} & \dy{1} && \nicefrac{1}{8} \\
      \dy{0} & \dy{1} && \dy{0} & \dy{0} && \dy{1} & \dy{0} && \nicefrac{1}{8} \\
      \dy{0} & \dy{1} && \dy{1} & \dy{0} && \dy{1} & \dy{1} && \nicefrac{1}{8} \\
      \dy{1} & \dy{0} && \dy{0} & \dy{1} && \dy{0} & \dy{0} && \nicefrac{1}{8} \\
      \dy{1} & \dy{0} && \dy{1} & \dy{1} && \dy{0} & \dy{1} && \nicefrac{1}{8} \\
      \dy{1} & \dy{1} && \dy{0} & \dy{1} && \dy{1} & \dy{0} && \nicefrac{1}{8} \\
      \dy{1} & \dy{1} && \dy{1} & \dy{1} && \dy{1} & \dy{1} && \nicefrac{1}{8} \\
      \bottomrule
    \end{tabular}
    \label{subtab:diadic}
  }
  \subfloat[Triadic]
  {
    \begin{tabular}{@{}cccccccccl@{}}
      \toprule
      \multicolumn{2}{c}{\var{X}} && \multicolumn{2}{c}{\var{Y}} && \multicolumn{2}{c}{\var{Z}} && \\
      \cmidrule(lr){1-2}  \cmidrule(lr){4-5}  \cmidrule(lr){7-8}
      \var{X_0} & \var{X_1} && \var{Y_0} & \var{Y_1} && \var{Z_0} & \var{Z_1} && $\Prob$ \\
      \midrule
      \tri{0} & \tri{0} && \tri{0} & \tri{0} && \tri{0} & \tri{0} && \nicefrac{1}{8} \\
      \tri{0} & \tri{1} && \tri{0} & \tri{1} && \tri{0} & \tri{1} && \nicefrac{1}{8} \\
      \tri{0} & \tri{0} && \tri{1} & \tri{0} && \tri{1} & \tri{0} && \nicefrac{1}{8} \\
      \tri{0} & \tri{1} && \tri{1} & \tri{1} && \tri{1} & \tri{1} && \nicefrac{1}{8} \\
      \tri{1} & \tri{0} && \tri{0} & \tri{0} && \tri{1} & \tri{0} && \nicefrac{1}{8} \\
      \tri{1} & \tri{1} && \tri{0} & \tri{1} && \tri{1} & \tri{1} && \nicefrac{1}{8} \\
      \tri{1} & \tri{0} && \tri{1} & \tri{0} && \tri{0} & \tri{0} && \nicefrac{1}{8} \\
      \tri{1} & \tri{1} && \tri{1} & \tri{1} && \tri{0} & \tri{1} && \nicefrac{1}{8} \\
      \bottomrule
    \end{tabular}
    \label{subtab:triadic_exp}
  }
  \caption{
    Expansion of the (a) dyadic and (b) triadic distributions. In both cases, the variables from Table\nobreakspace \ref {tab:distributions} were interpreted as two binary random variables, translating \eg $X = 3$ into $(X_0, X_1) = (1, 1)$. In this light, it becomes apparent that the dyadic distribution consists of the subdependencies $X_0 = Y_1$, $Y_0 = Z_1$, and $Z_0 = X_1$ while the triadic distribution consists of $X_0 + Y_0 + Z_0 = 0 \mod 2$ and $X_1 = Y_1 = Z_1$. These relationships are pictorially represented in Fig.\nobreakspace \ref {fig:dependencies}.
  }
  \label{tab:binary}
\end{table}

\begin{figure}
  \includegraphics{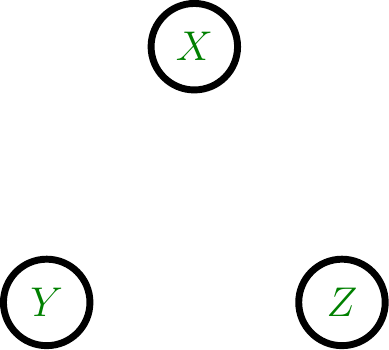}
  \caption{
    Graphical depiction of the result from applying several Bayesian network inference algorithms to both the dyadic and triadic distributions. The algorithms do not find any dependence between the variables \var{X}, \var{Y}, and \var{Z}, resulting in three disconnected nodes. The algorithms' failure is not surprising: the dyadic and triadic distributions cannot be represented by a directed acyclic graph, a basic assumption of Bayesian network inference.
  }
  \label{fig:bayesnet}
\end{figure}

Bayesian networks~\cite{jensen1996introduction} underlie common methods of dependency determination. And so, naturally, we applied Bayesian network inference to the two distributions; specifically, the \texttt{GS}, \texttt{IAMB}, \texttt{Fast-IAMB}, \texttt{Inter-IAMB}, \texttt{MMPC}, \texttt{SI-HITON-PC}, \texttt{HC}, \texttt{Tabu}, \texttt{MMHC}, and \texttt{RSMAX2} methods of the \texttt{bnlearn}~\cite{bnlearn} package, version 4.0. Each of these methods, when run with its default parameters, resulted in the Bayesian network depicted in Fig.\nobreakspace \ref {fig:bayesnet}: three lone nodes. Despite the popularity of Bayesian networks for modeling dependencies, the failure of these methods is not surprising: the dyadic and triadic distributions violate a basic premise of Bayesian networks. Namely, that their dependency structure cannot be represented by a directed acyclic graph. This also implies that the methods of Pearl~\cite{pearl2009causality} and its generalizations~\cite{ay2008information} give no insight on structure in these distributions. And so, we leave Bayesian network inference aside.

\begin{figure}
  \subfloat[Dyadic]
  {
    \includegraphics{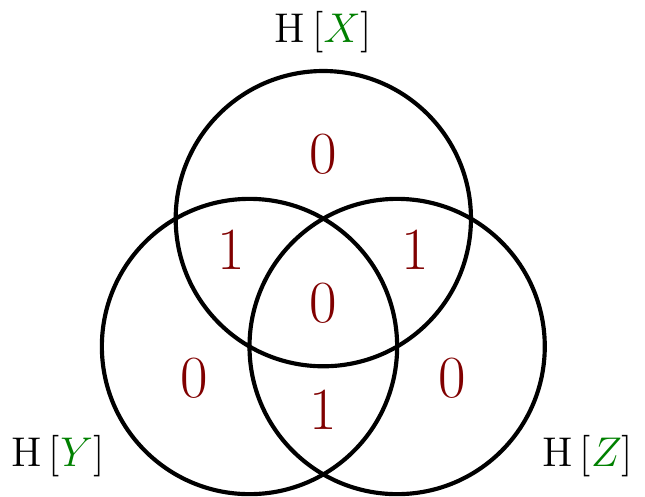}
    \label{subfig:dyadic_idiagram}
  }\\
  \subfloat[Triadic]
  {
    \includegraphics{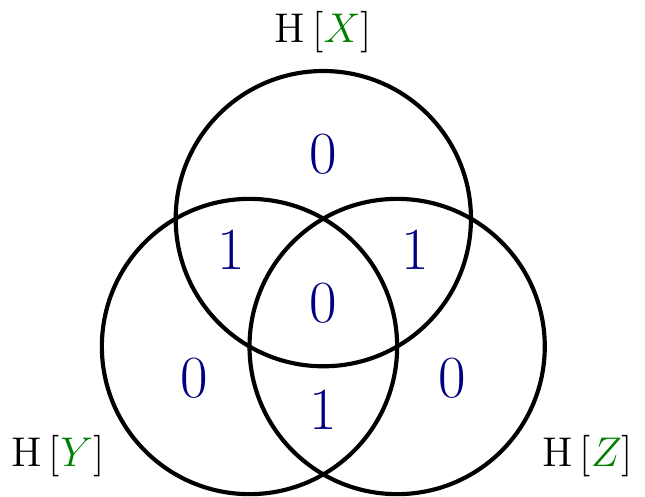}
    \label{subfig:triadic_idiagram}
  }
  \caption{
    Information diagrams for the (a) dyadic and (b) triadic distributions. For the three variable distributions depicted here, the diagram consists of seven atoms: three conditional entropies (each with value \SI{0}{\bit}), three conditional mutual informations (each with value \SI{1}{\bit}), and one co-information (\SI{0}{\bit}). Note that the two diagrams are \emph{identical}, meaning that although the two distributions are fundamentally distinct, no standard information-theoretic measure can differentiate the two.
  }
  \label{fig:idiagrams}
\end{figure}

What does an informational analysis say? Both the dyadic and triadic distributions describe events over three variables, each with an alphabet size of four. Each consists of eight joint events, each with a probability of \nicefrac{1}{8}. As such, each has a joint entropy of $\H{X, Y, Z} = \SI{3}{\bit}.$\footnote{The SI standard \emph{symbol} for information is \si{\bit}, analogous to \si{\second} being the symbol for time. As such it is inappropriate to write \SI{3}{\bit}s, just as it would be inappropriate to write \SI{3}{\second}s.} Our first observation is that \emph{any} entropy---conditional or not---and any mutual information---conditional or not---will be identical for the two distributions.  Specifically, the entropy of any variable conditioned on the other two vanishes: $\H{X \mid Y,Z} = \H{Y \mid X,Z} = \H{Z \mid X,Y} = \SI{0}{\bit}$; the mutual information between any two variables conditioned on the third is unity: $\I{X:Y \mid Z} = \I{X:Z \mid Y} = \I{Y:Z \mid X} = \SI{1}{\bit}$; and the three-way co-information also vanishes: $\I{X:Y:Z} = \SI{0}{\bit}$. These conclusions are compactly summarized in the form of the \emph{information diagrams} (I-diagrams)~\cite{reza1961introduction,yeung1991new} shown in Fig.\nobreakspace \ref {fig:idiagrams}. This diagrammatically represents all of the possible Shannon information measures (\emph{I-measures})~\cite{yeung1991new} of the distribution: effectively, all the multivariate extensions of the standard Shannon measures, called \emph{atoms}. The values of the information atoms are identical.

As a brief aside, it is of interest to note that it has been suggested (\eg, in Refs.~\cite{Bell03a,bettencourt2007functional}, among others) that zero co-information implies that at least one variable is independent of the others---that is, in this case, a lack of three-way interactions. Krippendorff~\cite{krippendorff2009information} early on demonstrated that this is not the case, though these examples more clearly exemplify this fact.

We now turn to the implications of the two information diagrams, Figs.\nobreakspace \ref {subfig:dyadic_idiagram} and\nobreakspace  \ref {subfig:triadic_idiagram}, being identical. There are measures~\cite{Bell03a,watanabe1960information,te1980multiple,chan2015multivariate,james2011anatomy,schreiber2000measuring,lamberti2004intensive,massey1990causality,marko1973bidirectional,sun2014causation} and expansions~\cite{bettencourt2008identification,bar2004multiscale,allen2014information} purporting to measure or otherwise extract the complexity, magnitude, or structure of dependencies within a multivariate distribution. Many of these techniques, including those just cited, are sums and differences of atoms in these information diagrams. As such, they are unable to differentiate these distributions.

To drive home the point that the concerns raised here are very broad, Table\nobreakspace \ref {tab:measures} enumerates the results of applying a great many information measures. It is organized in to four sections: entropies, mutual informations, common informations, and other measures.

None of the entropies, dependent only on the probability mass function of the distribution, can distinguish the two distributions. Nor can any of the mutual informations, as they are functions of the information atoms in the I-diagrams of Fig.\nobreakspace \ref {fig:idiagrams}.

The common informations, defined via auxiliary variables satisfying particular properties, can potentially isolate differences in the dependencies. Though only one of them---the G{\'a}cs-K{\"o}rner common information \K{\bullet}~\cite{gacs1973common,tyagi2011function}, involving the construction of the largest ``subrandom variable'' common to the variables and highlighted in the table---discerns that the two distributions are not equivalent because the triadic distribution contains the subrandom variable $X_1 = Y_1 = Z_1$ common to all three variables.

Finally, only two of the other measures (also highlighted) identify any difference between the two. Some fail because they are functions of the probability mass function. Others, like the TSE complexity~\cite{ay2006unifying} and erasure entropy~\cite{verdu2008information}, fail since they are functions of the I-diagram atoms. Only the intrinsic mutual information \I{\bullet\downarrow\bullet}~\cite{maurer1997intrinsic} and the reduced mutual information \I{\bullet\Downarrow\bullet}~\cite{renner2003new} distinguish the two since the dyadic distribution contains three dyadic subvariables each of which is independent of the third variable, whereas in the triadic distribution the conditional dependence of the \xor relation can be destroyed.

\def \gap {~~~~~~~~~~~}
\def \fnm {\footnotemark}
\begin{table}
  \centering
  \caption{Suite of information measures applied to the dyadic and triadic distributions, where:
  \H{\bullet} is the Shannon entropy~\cite{cover2012elements},
  \ReE{\bullet} is the R{\'e}nyi entropy~\cite{renyi1961measures},
  \TsE{\bullet} is the Tsallis entropy~\cite{tsallis1988possible},
  \I{\bullet} is the co-information~\cite{Bell03a},
  \T{\bullet} is the total correlation~\cite{watanabe1960information},
  \B{\bullet} is the dual total correlation~\cite{te1980multiple,abdallah2012measure},
  \J{\bullet} is the CAEKL mutual information~\cite{chan2015multivariate},
  \II{\bullet} is the interaction information~\cite{mcgill1954multivariate},
  \K{\bullet} is the G{\'a}cs-K{\"o}rner common information~\cite{gacs1973common},
  \C{\bullet} is the Wyner common information~\cite{wyner1975common,liu2010common},
  \G{\bullet} is the exact common information~\cite{kumar2014exact},
  \F{\bullet} is the functional common information\protect\fnm[1],
  \M{\bullet} is the MSS common information\protect\fnm[2],
  \I{\bullet\downarrow\bullet} is the intrinsic mutual information~\cite{maurer1997intrinsic}\protect\fnm[3],
  \I{\bullet\Downarrow\bullet} is the reduced intrinsic mutual information~\cite{renner2003new}\protect\fnm[3]\protect\fnm[4],
  \X{\bullet} is the extropy~\cite{lad2015extropy},
  \R{\bullet} is the residual entropy or erasure entropy~\cite{abdallah2012measure,verdu2008information},
  \P{\bullet} is the perplexity~\cite{jelinek1977perplexity},
  \D{\bullet} is the disequilibrium~\cite{lamberti2004intensive},
  \LMRP{\bullet} is the LMRP complexity~\cite{lamberti2004intensive}, and
  \TSE{\bullet} is the TSE complexity~\cite{ay2006unifying}.
  Only the G{\'a}cs-K{\"o}rner common information and the intrinsic mutual
  informations, highlighted, are able to distinguish the two distributions; the
  G{\'a}cs-K{\"o}rner common information via the construction of a subvariable
  ($\var{X_1} = \var{Y_1} = \var{Z_1}$) common to \var{X}, \var{Y}, and
  \var{Z}, and the intrinsic mutual informations via the relationship
  $\var{X_0} = \var{Y_1}$ being independent of \var{Z}.
  }
  \begin{tabular}{l@{\hspace{-1pt}}c@{\hspace{-1pt}}r@{\hspace{-1pt}}c@{\hspace{-1pt}}r}
    \toprule
    Measures                 &\gap& \dy{Dyadic}      &\gap& \tri{Triadic}    \\
    \midrule
    \H{\varsa}               &\gap& \SI{3}{\bit}     &\gap& \SI{3}{\bit}     \\
    \ReE[2]{\varsa}          &\gap& \SI{3}{\bit}     &\gap& \SI{3}{\bit}     \\
    \TsE[2]{\varsa}          &\gap& \SI{0.875}{\bit} &\gap& \SI{0.875}{\bit} \\
    \midrule
    \I{\varsb}               &\gap& \SI{0}{\bit}     &\gap& \SI{0}{\bit}     \\
    \T{\varsb}               &\gap& \SI{3}{\bit}     &\gap& \SI{3}{\bit}     \\
    \B{\varsb}               &\gap& \SI{3}{\bit}     &\gap& \SI{3}{\bit}     \\
    \J{\varsb}               &\gap& \SI{1.5}{\bit}   &\gap& \SI{1.5}{\bit}   \\
    \II{\varsb}              &\gap& \SI{0}{\bit}     &\gap& \SI{0}{\bit}     \\
    \midrule
    \rowcolor{rowcolor}
    \K{\varsb}               &\gap& \SI{0}{\bit}     &\gap& \SI{1}{\bit}     \\
    \C{\varsb}               &\gap& \SI{3}{\bit}     &\gap& \SI{3}{\bit}     \\
    \G{\varsb}               &\gap& \SI{3}{\bit}     &\gap& \SI{3}{\bit}     \\
    \F{\varsb}\fnm[1]        &\gap& \SI{3}{\bit}     &\gap& \SI{3}{\bit}     \\
    \M{\varsb}\fnm[2]        &\gap& \SI{3}{\bit}     &\gap& \SI{3}{\bit}     \\
    \midrule
    \rowcolor{rowcolor}
    \I{\varsc}\fnm[3]        &\gap& \SI{1}{\bit}     &\gap& \SI{0}{\bit}     \\
    \rowcolor{rowcolor}
    \I{\varsd}\fnm[3]\fnm[4] &\gap& \SI{1}{\bit}     &\gap& \SI{0}{\bit}     \\
    \X{\varsa}               &\gap& \SI{1.349}{\bit} &\gap& \SI{1.349}{\bit} \\
    \R{\varsb}               &\gap& \SI{0}{\bit}     &\gap& \SI{0}{\bit}     \\
    \P{\varsa}               &\gap& \SI{8}{}         &\gap& \SI{8}{}         \\
    \D{\varsa}               &\gap& \SI{0.761}{\bit} &\gap& \SI{0.761}{\bit} \\
    \LMRP{\varsa}            &\gap& \SI{0.381}{\bit} &\gap& \SI{0.381}{\bit} \\
    \TSE{\varsb}             &\gap& \SI{2}{\bit}     &\gap& \SI{2}{\bit}     \\
    \bottomrule
  \end{tabular}
  \label{tab:measures}
  \footnotetext[1]{$\displaystyle \F{\{X_i\}} = \min_{{\substack{\ind X_i | V \\ V = f(\{X_i\})}}} \H{V}$, where $\ind X_i | V$ means that the $X_i$ are conditionally independent given $V$.}
  \footnotetext[2]{$\displaystyle \M{\{X_i\}} = \H{\joinop (X_i \mss X_{\overline{i}})}$, where $X \mss Y$ is the minimal sufficient statistic~\cite{cover2012elements} of $X$ about $Y$ and $\joinop$ denotes the informational union of variables.}
  \footnotetext[3]{Though this measure is generically dependent on which variable(s) are chosen to be conditioned on, due to the symmetry of the dyadic and triadic distributions the values reported here are insensitive to permutations of the variables.}
  \footnotetext[4]{The original work~\cite{renner2003new} used the slightly more verbose notation \I{\bullet \downarrow\downarrow \bullet}.}
\end{table}

\begin{figure}
  \includegraphics{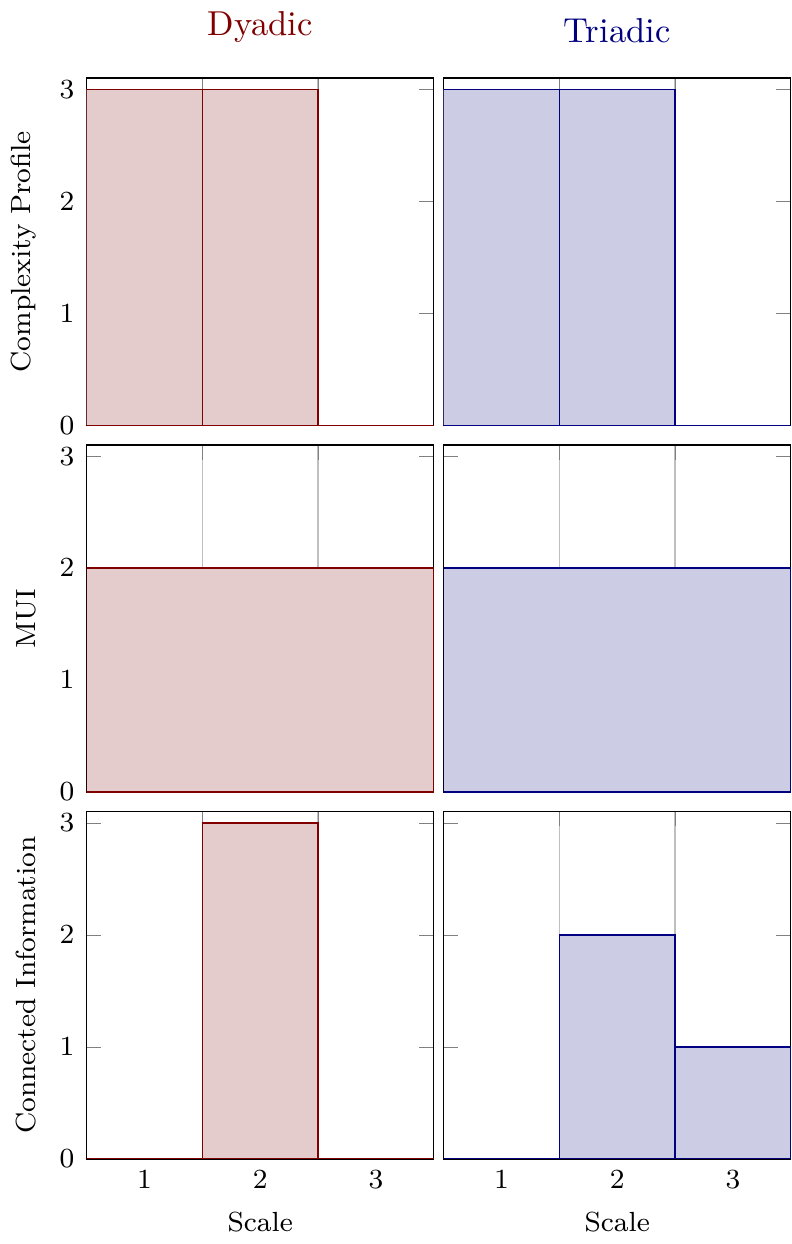}
  \caption{
    Suite of information expansions applied to the dyadic and triadic distributions: the complexity profile~\cite{bar2004multiscale}, the marginal utility of information~\cite{allen2014information}, and the connected information~\cite{schneidman2003network}. The complexity profile and marginal utility of information profiles are identical for the two distributions as a consequence of the information diagrams (Fig.\nobreakspace \ref {fig:idiagrams}) being identical. The connected informations, quantifying the amount of dependence realized by fixing $k$-way marginals, is able to distinguish the two distributions. Note that although all the x-axes are each scale, exactly what that means depends on the measure.
  }
  \label{fig:measures}
\end{figure}

Figure\nobreakspace \ref {fig:measures} demonstrates three different information expansions---that, roughly speaking, group variables into subsets of difference sizes or ``scales''---applied to our distributions of interest. The first is the complexity profile~\cite{bar2004multiscale}. At scale $k$, the complexity profile is the sum of all I-diagram atoms consisting of at least $k$ variables conditioned on the others. Here, since the I-diagrams are identical so are the complexity profiles. The second profile is the marginal utility of information~\cite{allen2014information}, which is a derivative of a linear programming problem whose constraints are given by the I-diagram so here, again, they are identical. Finally, we have Schneidman \etal's connected informations~\cite{schneidman2003network}, which are the differences in entropies of the maximum entropy distributions whose $k$- and $k-1$-way marginals are fixed to match those of the distribution of interest. Here, all dependencies are detected once pairwise marginals are fixed in the dyadic distribution, but it takes the full joint distribution to realize the \xor subdependency in the triadic distribution.

Neither the transfer entropy~\cite{schreiber2000measuring}, the transinformation~\cite{marko1973bidirectional}, the directed information~\cite{massey1990causality}, the causation entropy~\cite{sun2014causation}, nor any of their generalizations based on conditional mutual information are capable of determining that the dependency structure, and therefore the causal structure, of the two distributions are qualitatively different. This defect fundamentally precludes them from determining causal structure within a system of unknown dependencies. It underlies our prior criticism of these functions as measures of information flow~\cite{PhysRevLett.116.238701}.\footnote{As discussed there, the failure of these measures stems from the possibility of conditional dependence, whereas the aim for these directed measures is to quantify the information flow from one time series to another \emph{excluding} influences of the second. In this light, $\operatorname{T}_{X \to Y}^\prime = \I{X_0^t : Y_t \downarrow Y_0^t}$~\cite{maurer1997intrinsic} is certainly an incremental improvement over the transfer entropy.}

A promising approach to understanding informational dependencies is the \emph{partial information decomposition} (PID)~\cite{williams2010nonnegative}. This framework seeks to decompose a mutual information of the form \I{(I_0,I_1):O} into four nonnegative components: the information $R$ that both inputs $I_0$ and $I_1$ \emph{redundantly} provide the output $O$, the information $U_0$ that $I_0$ \emph{uniquely} provides $O$, the information $U_1$ that $I_1$ \emph{uniquely} provides $O$, and finally the information $S$ that both $I_0$ and $I_1$ \emph{synergistically} or \emph{collectively} provide $O$.

\begin{figure}
  \subfloat[Dyadic]
  {
    \includegraphics{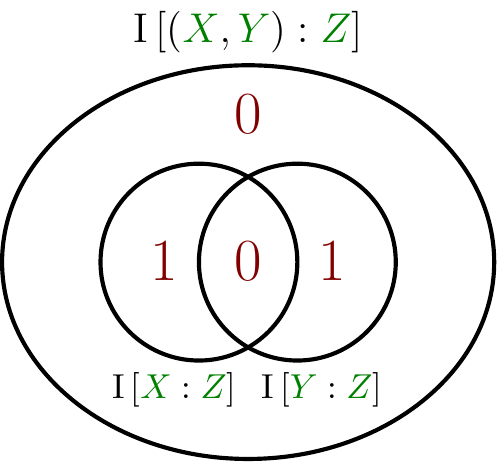}
    \label{subfig:dyadic_pid}
  }\\
  \subfloat[Triadic]
  {
    \includegraphics{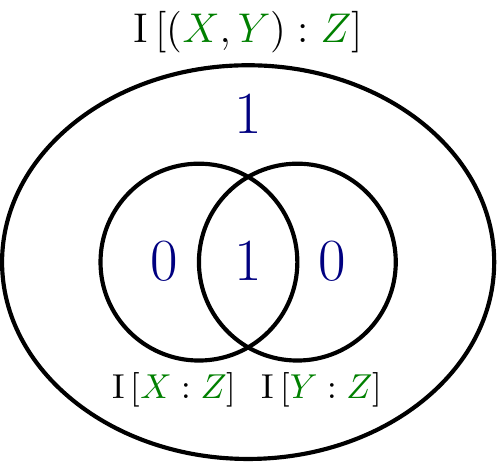}
    \label{subfig:triadic_pid}
  }
  \caption{
    Partial information decomposition diagrams for the (a) dyadic and (b) triadic distributions. Here, \var{X} and \var{Y} are treated as inputs and \var{Z} as output, but in both cases the decomposition is invariant to permutations of the variables. In the dyadic case, the relationship is realized as \SI{1}{\bit} of unique information from \var{X} to \var{Z} and \SI{1}{\bit} of unique information from \var{Y} to \var{Z}. In the triadic case, the relationship is quantified as \var{X} and \var{Y} providing \SI{1}{\bit} of redundant information about \var{Z} while also supplying \SI{1}{\bit} of information synergistically about \var{Z}.
  }
  \label{fig:pids}
\end{figure}

Under this decomposition, our two distributions take on very different characteristics.\footnote{Here, we quantified the partial information lattice using the now-commonly accepted technique of Ref.~\cite{bertschinger2014quantifying}, though calculations using two other techniques~\cite{harder2013bivariate,griffith2014intersection} match. The original PID measure $I_{\textrm{min}}$, however, assigns both distributions \SI{1}{\bit} of redundant information and \SI{1}{\bit} of synergistic information. We have yet to compute two other recent proposals~\cite{quax2016stripping,ince2016measuring}, though we suspect they will match Ref.~\cite{bertschinger2014quantifying}'s values.} For both, the decomposition is invariant as far as which variables are selected as $I_0$, $I_1$, and $O$. For the dyadic distribution PID identifies both bits in \I{(I_0,I_1):O} as unique, one from each input $I_i$, corresponding to the dyadic subdependency shared by $I_i$ and $O$. Orthogonally, for the triadic distribution PID identifies one of the bits as redundant, stemming from $X_1 = Y_1 = Z_1$, and the other as synergistic, resulting from the \xor relation among $X_0$, $Y_0$ and $Z_0$. These decompositions are displayed pictorially in Fig.\nobreakspace \ref {fig:pids}.

Another somewhat similar approach is that of \emph{integrated information theory}~\cite{iit3}. However, this approach requires a known dynamic over the variables and is, in addition, highly sensitive to the dynamic. Here, in contrast, we considered only simple probability distributions without any assumptions as to how they might arise from the dynamics of interacting agents. That said, one might associate an integrated information measure with a distribution via the maximal information integration over all possible dynamics that give rise to the distribution. We leave this task for a later study.

\section{Discussion}
\label{sec:discussion}

The broad failure of Shannon information measures to differentiate the dyadic and polyadic distributions has far-reaching consequences. Consider, for example, an experiment where a practitioner places three probes into a cluster of neurons, each probe touching two neurons and reporting 0 when they are both quiescent, 1 when the first is excited but second quiescent, 2 when the second is excited but the first quiescent, and 3 when both are excited. Shannon-like measures---including the transfer entropy---would be unable to differentiate between the dyadic situation consisting of three pairs of synchronized neurons, the triadic situation consisting of a trio of synchronized neurons, and a trio exhibiting the \xor relation---a relation requiring nontrivial sensory integration. Such a situation might arise when probing the circuitry of the \emph{drosophila melanogaster} connectome~\cite{takemura2013visual}, for instance.

Furthermore, while partitioning each variable into subvariables made the dependency structure clear, we do not believe that such a refinement should be a necessary step in discovering such structure. Consider that we demonstrated that refinement is not strictly needed, since the partial information decomposition was able to discover the distribution's internal structure without it.

These results, observations, and the broad survey clearly highlight the need to extend Shannon's theory. In particular, the extension must introduce a fundamentally new measure, not merely sums and differences of the standard Shannon information measures. While the partial information decomposition was initially proposed to overcome the interpretational difficulty of the (potentially negative valued) co-information, we see here that it actually overcomes a vastly more fundamental weakness with Shannon information measures. While negative information atoms can \emph{subjectively} be seen as a flaw, the inability to distinguish dyadic from polyadic relations is a much deeper and objective issue.

This may lead one to consider the partial information decomposition as the needed extension to Shannon theory. We do not. The partial information decomposition depends on interpreting some random variables as ``inputs'' and others as ``outputs''. While this may be perfectly natural in some contexts, it is not satisfactory in general. Consider: how should the triadic distribution's information be allotted? Certainly one of its three bits is redundant and one is synergistic, but what about the third? The \xor (dyadic) distribution contains two bits---are both to be considered synergy~\cite{e18020038}? In any case, the partial information framework does not address this question.

\begin{figure}
  \subfloat[Distribution]{
    \begin{minipage}[c]{0.275\columnwidth}
      \begin{tabular}{@{}ccccl@{}}
        \toprule
        \var{W} & \var{X} & \var{Y} & \var{Z} & $\Prob$ \\
        \midrule
        0 & 0 & 0 & 0 & \nicefrac{1}{8} \\
        0 & 1 & 3 & 1 & \nicefrac{1}{8} \\
        1 & 0 & 2 & 2 & \nicefrac{1}{8} \\
        1 & 1 & 1 & 3 & \nicefrac{1}{8} \\
        2 & 2 & 3 & 3 & \nicefrac{1}{8} \\
        2 & 3 & 0 & 2 & \nicefrac{1}{8} \\
        3 & 2 & 1 & 1 & \nicefrac{1}{8} \\
        3 & 3 & 2 & 0 & \nicefrac{1}{8} \\
        \bottomrule
      \end{tabular}
    \end{minipage}
    \label{subtab:distribution_camo}
  }%
  \subfloat[I-diagram]{
    \begin{minipage}[c]{0.65\columnwidth}
      \includegraphics{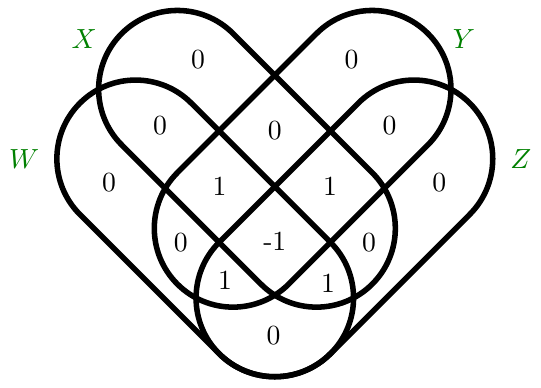}
    \end{minipage}
    \label{subfig:idiagram_camo}
  }
  \caption{
    Dyadic camouflage distribution: This distribution, when uniformly and independently mixed with the 4-variable parity distribution (in which each variable is the parity of the other three), results in a distribution whose I-diagram incorrectly implies that the distribution contains only dyadic dependencies.
  }
  \label{fig:camo}
\end{figure}

\section{Dyadic Camouflage \& Dependency Diffusion}
\label{sec:camouflage}

The dyadic and triadic distributions we analyzed thus far were deliberately chosen to have small dimensionality in an effort to make them and the failure of Shannon information measures as comprehensible and intuitive as possible. Since a given data set may have exponentially many different three-variable subsets, even just this pair of trivariate distributions will stymie most any assessment of variable dependency. However, this is simply a starting point. We will now demonstrate that there exist distributions of arbitrary size which mask their $k$-way dependencies in such a way that they match, from a Shannon information theory prospective, a distribution of the same size containing only dyadic relationships.  Furthermore, we show how any such distribution may be obfuscated over any larger set of variables. This likely mandates a search over all partitions of all subsets a system, making the problem of finding such distributions in the EXPTIME computational complexity class \cite{Gare79}.

Specifically, consider the $4$-variable parity distribution consisting of four binary variables such that each variable's value is equal to the parity of the remaining three. This is a straightforward generalization of the \xor distribution used in constructing the triadic distribution. We next need a generalization of the ``giant bit''~\cite{abdallah2012measure}---which we call \emph{dyadic camouflage}---to mix with the parity, informationally ``canceling out'' the higher-order mutual informations even though dependencies of such orders exist in the distribution. An example dyadic camouflage distribution for four variables is given in Fig.\nobreakspace \ref {fig:camo}.

Generically, an $n$-variable dyadic camouflage distribution has an alphabet size of $2^{n-2}$ and consists of $2^{\frac{(n-2)\cdot(n-1)}{2}}$ equally likely outcomes, both numbers determined due to entropy considerations. The distribution is constrained such that any two variables are completely determined by the remaining $n-2$. Moreover, each $m$-variable subdistribution has equal entropy and, otherwise, is of maximal joint entropy. One method of constructing such a distribution is to begin by writing down one variable in increasing lexicographic order such that it has the correct number of outcomes; \eg, column $W$ in Fig.\nobreakspace \ref {subtab:distribution_camo}. For each remaining variable, uniformly apply permutations of increasing size to its set of values.

Finally, one can obfuscate any distribution by embedding it in a larger collection of random variables. Given a distribution $D$ over $n$ variables, associate each random variable $i$ of $D$ with a $k$-variable subset of a distribution $D^\prime$ in such a way that there is a mapping from the $k$ outcomes in the subset of $D^\prime$ to the outcome of the variable $i$ in $D$. For example, one can embed the \xor distribution over $X, Y, Z$ into six variables $X_0, X_1, Y_0, Y_1, Z_0, Z_1$ via $X_0 \oplus X_1 = X$, $Y_0 \oplus Y_1 = Y$, and $Z_0 \oplus Z_1 = Z$. In other words, the parity of $(Z_0, Z_1)$ is equal to the \xor of the parities of $(X_0, X_1)$ and $(Y_0, Y_1)$. In this way one must potentially search over all partitions of all subsets of $D^\prime$ in order to discover the distribution $D$ hiding within. We refer to this method of obfuscation as \emph{dependency diffusion}.

The first conclusion is that the challenges of conditional dependence can be found in joint distributions over arbitrarily large sets of random variables. The second conclusion, one that heightens the challenge to discovery, is that even finding which variables are implicated in polyadic dependencies can be exponentially difficult. Together the camouflage and diffusion constructions demonstrate how challenging it is discover, let alone work with, multivariate dependencies. This difficulty strongly implies that the current state of information-theoretic tools is vastly underpowered for the types of analyses required of our modern, data-rich sciences.

It is unlikely that the parity plus dyadic camouflage distribution discussed here is the only example of Shannon measures conflating the arity of dependencies and thus producing an information diagram identical to that of a qualitatively distinct distribution. This suggests an important challenge: find additional, perhaps simpler, joint distributions exhibiting this phenomenon.

\section{Conclusion}
\label{sec:conclusion}

To conclude, we constructed two distributions that cannot be distinguished using conventional (and many nonconventional) Shannon-like information measures. In fact, of the more than two dozen measures we surveyed only five were able to separate the distributions: the G{\'a}cs-K{\"o}rner common information, the intrinsic mutual information, the reduced intrinsic mutual information, the connected informations, and the partial information decomposition. We also noted in an aside that causality detection approaches that assume an underlying directed acyclic graph structure are structurally impotent.

The failure of the Shannon-type measures is perhaps not surprising: nothing in the standard mathematical theories of information and communication suggests that such measures \emph{should} be able to distinguish these distributions \cite{Shan56b}. However, distinguishing dyadic from triadic relationships and the related causal structure is of the utmost importance to the sciences. Critically, since interpreting dependencies in random distributions is traditionally the domain of information theory, we propose that new extensions to information theory are needed.

Furthermore, the dyadic camouflage distribution (Fig.\nobreakspace \ref {fig:camo}) presents an acute challenge for traditional methods of dependency and causality inference. Let's close with an example. Consider the widely used Granger causality~\cite{granger1969investigating} applied to the camouflage distribution. Fixing any two variables, say $X$ and $Y$, determines the remaining two, $Z$ and $W$. What is one to conclude from this, other than that $X$ cannot influence $Y$? And yet, in conjunction with either $Z$ or $W$, $X$ completely determines $Y$. This makes clear the deep assumption of dyadic relationships that permeates and biases our ways of thinking about complex systems.

These results may seem like a deal-breaking criticism of employing information theory to determine dependencies. Indeed, these results seem to indicate that much existing empirical work and many interpretations have simply been wrong and, worse even, that the associated methods are misleading while appearing quantitatively consistent. We think not, though. With the constructive and detailed problem diagnosis given here, at least we can see the true problem. It is now a necessary step to address it. This leads us to close with a cautionary quote:

\epigraph{\emph{The tools we use have a profound (and devious!) influence on our thinking habits, and, therefore, on our thinking abilities.}}{Edsger W. Dijkstra~\cite{dijkstra1982we}}

\section*{Acknowledgments}
\label{sec:acknowledgments}

We thank N. Barnett and D. P. Varn for helpful feedback. JPC thanks the Santa Fe Institute for its hospitality during visits as an External Faculty member. This material is based upon work supported by, or in part by, the U. S. Army Research Laboratory and the U. S. Army Research Office under contracts W911NF-13-1-0390 and W911NF-13-1-0340.

\appendix

\section{A Python Discrete Information Package}

Hand calculating the information quantities used in the main text, while profitably done for a few basic examples, soon becomes tedious and error prone. We provide a Jupyter notebook~\cite{jupyter} making use of \texttt{dit} (``Discrete Information Theory'')~\cite{dit}, an open source Python package that readily calculates these quantities.

\end{document}